\documentclass[%
 reprint,
superscriptaddress,
 amsmath,amssymb,
 aps,
floatfix,
prd,
]{revtex4-2}

\def\apjs{Astrophysical Journal, Supplement} 
\def\sovast{Soviet Astronomy}      
\def\azh{Astronomicheskii Zhurnal}    

\usepackage{graphicx}
\usepackage{dcolumn}
\usepackage{bm}
\usepackage{hyperref}
\usepackage[dvipsnames]{xcolor}
\usepackage{comment}
\usepackage{orcidlink}

\definecolor{lightsabergreen}{rgb}{.14,.64,.14}
\definecolor{lightgreen}{rgb}{.14,.44,.14}

\hypersetup{
  colorlinks   = true, 
  urlcolor     = Maroon, 
  linkcolor    = Maroon, 
  citecolor   = Maroon 
}

\begin{document}


\title{Dissipative Dark Cosmology: \\From Early Matter Dominance to Delayed Compact Objects}

\author{Joseph Bramante\orcidlink{0000-0001-8905-1960}}
\email{joseph.bramante@queensu.ca}
\affiliation{Department of Physics, Engineering Physics, and Astronomy,
Queen’s University, Kingston, Ontario, K7L 2S8, Canada}
\affiliation{Arthur B. McDonald Canadian Astroparticle Physics Research Institute, Kingston ON K7L 3N6, Canada}
\affiliation{Perimeter Institute for Theoretical Physics, Waterloo, ON N2J 2W9, Canada}

\author{Christopher V. Cappiello\orcidlink{0000-0002-7466-9634}}
\email{cvc1@queensu.ca}
\affiliation{Department of Physics, Engineering Physics, and Astronomy,
Queen’s University, Kingston, Ontario, K7L 2S8, Canada}
\affiliation{Arthur B. McDonald Canadian Astroparticle Physics Research Institute, Kingston ON K7L 3N6, Canada}
\affiliation{Perimeter Institute for Theoretical Physics, Waterloo, ON N2J 2W9, Canada}
\affiliation{Department of Physics, Washington University, St. Louis, MO, 63130, USA}

\author{Melissa Diamond\orcidlink{0000-0003-1221-9475}}%
\email{m.diamond@queensu.ca}
\affiliation{Department of Physics, Engineering Physics, and Astronomy,
Queen’s University, Kingston, Ontario, K7L 2S8, Canada}
\affiliation{Arthur B. McDonald Canadian Astroparticle Physics Research Institute, Kingston ON K7L 3N6, Canada}
\affiliation{Perimeter Institute for Theoretical Physics, Waterloo, ON N2J 2W9, Canada}

\author{J. Leo Kim\orcidlink{0000-0001-8699-834X}}%
\email{leo.kim@queensu.ca}
\affiliation{Department of Physics, Engineering Physics, and Astronomy,
Queen’s University, Kingston, Ontario, K7L 2S8, Canada}
\affiliation{Arthur B. McDonald Canadian Astroparticle Physics Research Institute, Kingston ON K7L 3N6, Canada}

\author{Qinrui Liu\orcidlink{0000-0003-3379-6423}}%
\email{qinrui.liu@queensu.ca}
\affiliation{Department of Physics, Engineering Physics, and Astronomy,
Queen’s University, Kingston, Ontario, K7L 2S8, Canada}
\affiliation{Arthur B. McDonald Canadian Astroparticle Physics Research Institute, Kingston ON K7L 3N6, Canada}
\affiliation{Perimeter Institute for Theoretical Physics, Waterloo, ON N2J 2W9, Canada}

\author{Aaron C. Vincent\orcidlink{0000-0003-3872-0743}}%
\email{aaron.vincent@queensu.ca}
\affiliation{Department of Physics, Engineering Physics, and Astronomy,
Queen’s University, Kingston, Ontario, K7L 2S8, Canada}
\affiliation{Arthur B. McDonald Canadian Astroparticle Physics Research Institute, Kingston ON K7L 3N6, Canada}
\affiliation{Perimeter Institute for Theoretical Physics, Waterloo, ON N2J 2W9, Canada}

\date{\today}

\begin{abstract}
We demonstrate a novel mechanism for producing dark compact objects and black holes through a dark sector, where all the dark matter can be dissipative. Heavy dark sector particles with masses above $10^4$ GeV can come to dominate the Universe and yield an early matter-dominated era before Big Bang Nucleosynthesis (BBN). Density perturbations in this epoch can grow and collapse into tiny dark matter halos, which cool via self interactions. The typical halo size is set by the Hubble length once perturbations begin growing, offering a straightforward prediction of the halo size and evolution depending on ones choice of dark matter model. Once these primordial halos have formed, a thermal phase transition can then shift the Universe back into radiation domination and standard cosmology. These halos can continue to collapse after BBN, resulting in the late-time formation of fragmented dark compact objects and sub-solar mass primordial black holes. We find that these compact objects can constitute a sizable fraction of all of dark matter. The resulting fragments can have masses between $10^{20}$ g to $10^{32}$ g, with radii ranging from $10^{-2}$ m to $10^5$ m, while the black holes can have masses between $10^{8}$ g to $10^{34}$ g. Furthermore, a unique feature of this model is the late-time formation of black holes which can evaporate today. We compare where these objects lie with respect to current primordial black hole and and massive (astrophysical) compact halo object constraints.
\end{abstract}

\maketitle

\section{Introduction}

Cold, collisionless dark matter (DM) works well to explain structure formation over a wide range of scales. However, dark matter's non-gravitational interactions, especially its possible self interactions, are poorly understood. Given the complexity and variety of forces within the Standard Model (SM), it is worth considering similar interactions within the dark sector. In fact, DM self interactions have previously been invoked to explain  potential problems in cosmology, such as issues with small scale structure formation~\cite{Kaplan_2010,Cyr_Racine_2013,Boddy_2016,Foot_2015,Foot_2016} and the Hubble tension~\cite{bansal2023precision}. More generally, related studies have explored how such a dissipative dark sector could produce collapsed structures like dark stars and black holes, though typically this dissipative sector is assumed to be only a subdominant component of the total DM budget~\cite{Fan:2013tia, Fan:2013yva, Kouvaris:2015rea, Foot:2016wvj, Buckley:2017ttd, DAmico:2017lqj, Chang:2018bgx, Latif:2018kqv, Gurian:2022nbx, Fernandez:2022zmc, Roy:2023zar, Bramante:2023ddr,Gemmell:2023trd}.

Recently, several works have studied dark structure formation prior to big bang nucleosynthesis (BBN). While some have explored the behavior of dissipative dark sectors assuming radiation domination at these early times \cite{Amendola:2017xhl, Savastano:2019zpr, Flores:2020drq, Flores:2023nto, Flores:2023zpf}, others have indicated that collisionless dark matter can form structures in an early matter-dominated era (EMDE) \cite{Delos:2019mxl, Blanco:2019eij, Ganjoo:2022rhk, Ganjoo:2023fgg}.  While EMDEs have been getting more attention lately, they have a long history within cosmology theory, reaching back to early work on primordial black holes (PBHs) \cite{Hawking:1971ei, Carr:1974nx}.  PBHs can come to dominate the Universe and evaporate away before BBN \cite{Carr:1976zz, Anantua:2008am} or form from structure collapse during an EMDE and persist to late times \cite{Khlopov:1980mg, Polnarev_1981, Polnarev_1982}.

In this work, we present a new cosmological pathway for the formation of heavy dark compact objects and black holes. We show that a dissipative dark sector can collapse and rapidly cool in an EMDE prior to BBN. We consider a remarkably simple dark sector with only two constituents -- a dark Dirac fermion and a massive dark vector mediator. We consider a wide range of particle masses, varying from $10^4$ GeV up to $10^{16}$ GeV. During the EMDE, density perturbations grow and eventually collapse into virialized, primordial dark matter halos. These cool slowly over time, and ultimately either collapse into black holes or fragment into dark, compact, pressure supported objects, henceforth called DarkCOs. While an EMDE would usually result in a universe perpetually dominated by dark matter, a short period of thermal inflation, we will show, can dilute the heavy states and return the universe to radiation domination prior to BBN. We show this using a model where a single heavy scalar field coupled to the SM thermal bath drives thermal inflation before decaying into SM radiation. This is similar to some previous proposals considering the effect of moduli fields on SUSY sectors \cite{Lyth:1995hj, Lyth:1995ka} as well as the same effect on generalized WIMP dark matter \cite{Davoudiasl:2015vba,Randall:2015xza, Berlin:2016gtr, Bramante:2017obj, Evans:2019jcs}. We model compact object formation for a wide range of model parameters, and find that this model produces a diverse population of black holes and fragmented, dark, pressure-supported compact objects. 

This paper is structured as follows. Section \ref{sec:perturbations} describes the growth and evolution of of dark density perturbations. Section \ref{sec:final_masses} describes the primordial dark halo cooling and collapse process. Section \ref{sec:inflation} discusses the transition from the EMDE to a radiation-dominated era via thermal inflation. In section \ref{sec:pheno}, we discuss some of the phenomenology of the black holes and compact objects formed, including existing constraints from evaporation, gravitational waves and gravitational lensing. We also note that these dark sector models are allowed by current self interacting dark matter (SIDM) constraints from the bullet cluster. We conclude in section \ref{sec:conclusion}.

\section{Dark sector overdensity and growth}\label{sec:perturbations}

We consider a simple, asymmetric dark sector comprised of a Dirac fermion $X$ which we call the \emph{dark electron}, with mass $m_X$, and a massive vector mediator $A_\mu$ which we call the \emph{dark photon} ($\gamma_D$), with mass $m_{\gamma_D}$. The interaction Lagrangian is given by 
\begin{equation}
    \mathcal{L} \supset \bar{X} (i \gamma^\mu D_\mu - m_X) X  - \frac{1}{4} F_{\mu \nu} F^{\mu \nu} + \frac{1}{2}m_{\gamma_D}^2 A_\mu A^\mu,
\end{equation}
where $D_\mu = \partial_\mu - i g_D A_\mu$ is the gauge covariant derivative with $\alpha_D = g_D^2/(4\pi)$.
In this work, we will focus on the regime where $1/m_{\gamma_D}$ is small compared to the length scale of all perturbations or compact objects considered.

\begin{figure*}
    \centering
    \includegraphics[width=.75\textwidth]{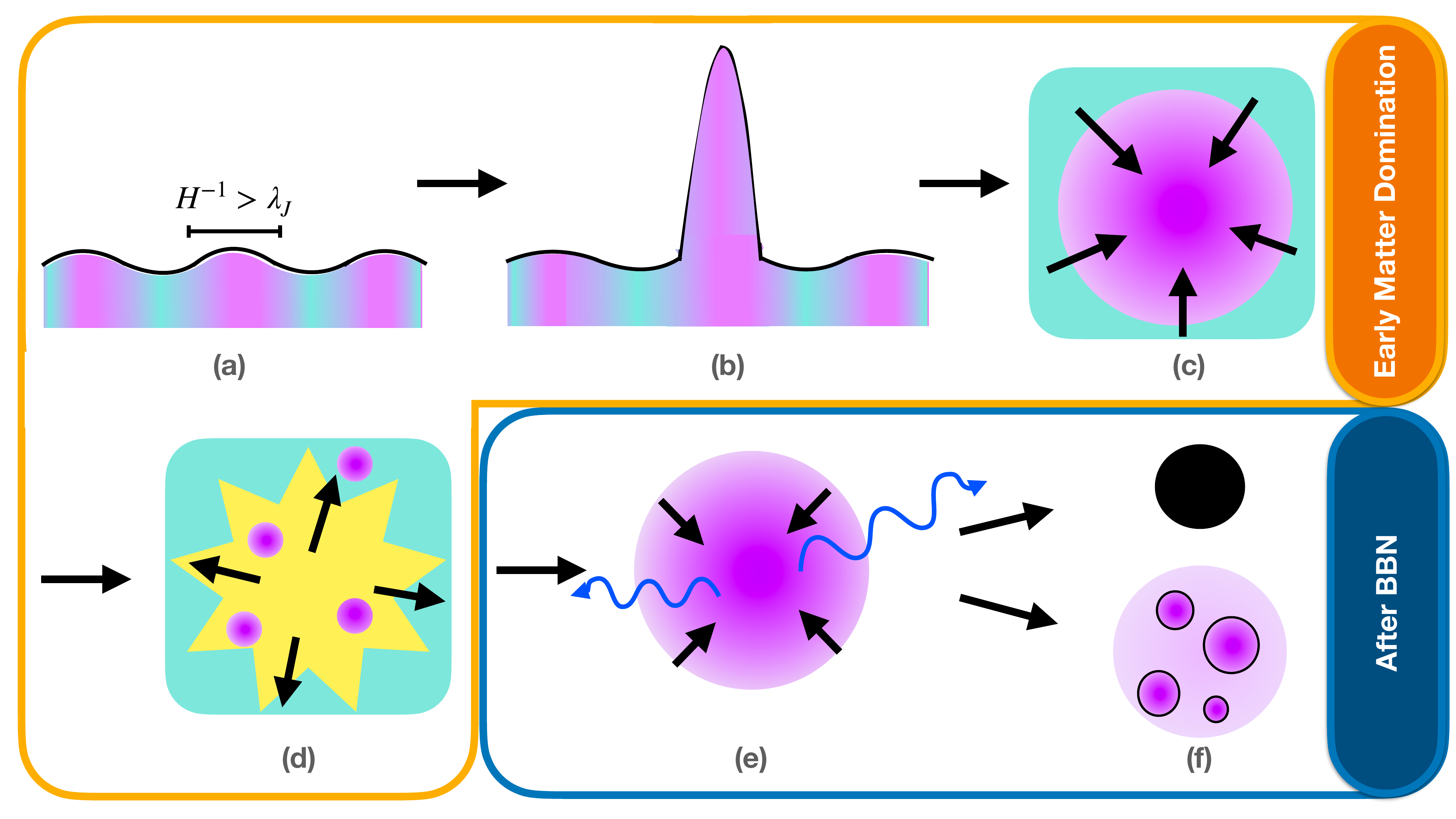}
    \caption{Evolution of dark electron structures from initial density perturbations to final collapsed form.  
    \textbf{(a)} During the EMDE dark electron density perturbations begin growing once the Hubble length exceeds the Jean's length, $\lambda_J$.
    \textbf{(b)} Overdensities grow linearly with the scale factor.
    \textbf{(c)} Once density perturbations grow to $\delta\sim 1$, the overdense patches decouple from the background expansion, begin self gravitating, and virialize into primordial dark electron halos.
    \textbf{(d)} Thermal inflation ends the EMDE, diluting the dark electron abundance without disrupting virialized, primordial dark electron halos, and initiating a SM radiation dominated era.
    \textbf{(e)} The primordial dark electron halos cool via dark bremsstrahlung.
    \textbf{(f)} When cooling is fast compared to collapse, the dark electron halos fragment into pressure supported structures which we call DarkCOs.  If cooling is slow compared to collapse the halo collapses directly to a black hole. }
    \label{cartoon}
\end{figure*}

Following a standard inflationary epoch and a subsequent period of reheating, we assume the SM and the dark sector are in thermal equilibrium in a relativistic bath of particles. Once the temperature of the particle bath drops below the dark matter mass, its abundance becomes Boltzmann-suppressed. In the presence of a large enough initial asymmetry (e.g. an $\mathcal{O}(1)$ asymmetry  as expected in Affleck-Dine models \cite{Affleck:1984fy,Bramante:2017obj}), then the comoving number density will rapidly freeze-out, and proceed to become the dominant energy density of the Universe as the relativistic plasma is redshifted away, leading to an era of early matter domination. We note that in this work the dark electrons will also come to constitute most of the dark matter, since although $\gamma_D$ is also a potential dark matter component, we will only consider $m_{\gamma_D} \ll m_X$, meaning the dark photon will redshift like matter at a lower temperature than the dark electron, from which we expect the relative dark photon$/$electron late-time abundance to be $\sim m_{\gamma_D} / m_X $.

The growth of perturbations and the collapse of overdensities proceed in an $X$-dominated EMDE as they would in an ordinary SM matter dominated era. We assume that at the start of the EMDE, the initial density perturbations, $\delta = (\rho-\bar{\rho})/\bar{\rho}$, are of order $\delta_0\sim 10^{-5}$, similar to typical primordial perturbations observed at late times \cite{Workman:2022ynf}.  We note that one may choose different, potentially larger, initial density perturbations for these small modes entering the horizon at such early times as in some treatments of EMDEs \cite{Delos:2019mxl,Khlopov:1980mg}. Perturbations grow linearly with the scale factor $a$, until turnaround, when $\delta\sim 1$, at which point the perturbations begin to collapse, and virialize independently of the background expansion of the Universe.  A cartoon depiction of the full evolution of dark electron structures from density perturbation to final collapsed object can be found in Fig. \ref{cartoon}. 

Only subhorizon perturbations that satisfy the Jeans criterion $\lambda \geq \lambda_J$ grow linearly during matter domination.  Here where $\lambda$ is the size of the perturbation and $\lambda_J$ is the Jeans length, given by
\begin{align}
    \lambda_J = c_s \left( \frac{\pi}{\rho_X G} \right)^{1/2},
\end{align}
where $\rho_X$ is the energy density of the perturbation. $c_s$ is the sound speed of the fluid, which can be obtained through its pressure \cite{Kouvaris:2015rea, Chang:2018bgx}
\begin{align}
\label{eq:pressure}
    P_X = n_X T_X + \frac{2 \pi \alpha_D n_X^2}{m_{\gamma_D}^2} ,
\end{align}
via the relationship $c_s^2 = dP_X/d\rho_X$. $n_X = \rho_X/m_X$ and $T_X$ are the number density and temperature of the dark electron gas, respectively. The first term in the pressure corresponds to the kinetic pressure of the $X$ particles, while the second term is due to the dark photon repulsive pressure.

The resulting sound speed then, is given by
\begin{align}
    c_s^2 = \frac{T_X}{m_X} + \frac{4 \pi \alpha_D n_X }{m_X m_{\gamma_D}^2}. \label{eq:cs_sq}
\end{align}
Note that while Fermi pressure from the dark electrons contributes to the overall pressure, it is subdominant to the dark photon repulsive pressure in all of the parameter space we consider.

Once $\lambda_J$ falls below the Hubble length, perturbations can begin to grow.  As the first perturbations to begin growing would be those entering the horizon just as $1/H\sim\lambda_J$, the typical sizes of perturbations---and the halos and compact objects they eventually form---are set by the horizon size when $1/H\sim\lambda_J$. We denote the temperature of the SM thermal bath at this moment as $T_\mathrm{grow}$. The associated mass of the growing perturbation is its horizon mass, $M_H$, $i.e.$ the total mass inside of the Hubble radius $H_\mathrm{grow}$ at growth, given by $M_H = 4\pi\rho_\mathrm{grow} H_\mathrm{grow}^{-3}/3$, where $\rho_\mathrm{grow} = 3 \zeta(3) m_X T_\mathrm{grow}^3/(4\pi^2)$ and $H_\mathrm{grow} = (8\pi G \rho_\mathrm{grow} /3)^{1/2}$.  $T_\mathrm{grow}$ is given by
\begin{equation}
    T_\mathrm{grow}= \text{min}\left[\sqrt{\frac{3}{80 \pi^2}}m_X,\left(\frac{m_X m_{\gamma_D}^2}{2 \pi \zeta(3) \alpha_D}\right)^{1/3}\right], \label{eq:T_grow}
\end{equation}
where the two arguments are for the cases where kinetic pressure and dark photon repulsive pressure dominate the sound speed in Eq. \eqref{eq:cs_sq}, respectively.  After growing linearly until $\delta\sim 1$, these overdense perturbations will collapse into a dark electron halo and virialize.

The dark halo can only cool efficiently throughout all stages of its collapse if it becomes hot enough, $T_X>m_{\gamma_D}$, to emit dark photons during virialization. In order for the gas to heat up during virialization, the dark electrons must be collisional with themselves during infall.  This is guaranteed if the dark halo is collisional at its maximum radius, $r_\mathrm{ta} = (8\pi G\rho_\mathrm{grow}/3)^{-1/2}  \delta_0^{-1}$ ($i.e.$ at turnaround), since the scattering length $\ell_{X}$ will decrease faster than the radius of the newly formed halo. The relevant differential cross section for $X+X$ scattering is 
\begin{equation}
    \frac{d \sigma}{d \Omega} = \frac{m_{X}^2\alpha_D^2(4p^4(1+3\cos^2\theta)+4p^2m_{\gamma_D}^2+m_{\gamma_D}^4)}{(4p^4\sin^2\theta+4p^2m_{\gamma_D}^2+m_{\gamma_D}^4)^2}\,.
\end{equation}
Here $p$ is the momentum of the dark electrons in the center of momentum frame, and $\theta$ is the scattering angle. We take $p^2\sim 3 m_X T_X$. Note that in the limit of a massless mediator, this reduces to the familiar cross section for M\o ller scattering. We estimate whether the gas is collisional by requiring $N_{X, \mathrm{sc}}\ell_{X} < r_\mathrm{ta}$, where $N_{X, \mathrm{sc}}$ is the number of scatters required for the $X$ particle to exchange half of its kinetic energy while crossing the halo, given by
\begin{equation}
   N_{X, \mathrm{sc}} = \frac{0.5\int d\cos\theta \frac{d\sigma}{d\Omega}}{\int d\cos\theta \frac{d\sigma}{d\Omega}(1-\cos(\theta))} ,
\end{equation}
and $\ell_{X}$ is the scattering length, given by 
\begin{align}
    \ell_{X} = \left( n_X \int \frac{d\sigma}{d\Omega} d\Omega\right)^{-1}.
\end{align}

\section{Dark sector collapse and final masses}
\label{sec:final_masses}
The primordial dark halo cools via bremsstrahlung, with a cooling rate given by \cite{Haug:1975, Chang:2018bgx}
\begin{align}
    \Lambda = \frac{32 \alpha_D^3 \rho_{X} T_{X}}{\sqrt{\pi} m_{X}^3 } \sqrt{\frac{T_{X}}{m_{X}}} e^{-m_{\gamma_D}/T_{X}} e^{-V^{1/3} \sqrt{N_{\mathrm{sc}}} / \ell_{\gamma_D}^{\mathrm{abs}} } ,
\end{align}
where the final exponential factor is to account for dark photon reabsorption. The factor $N_\mathrm{sc} = (V^{1/3} /\ell_{\gamma_D}^C )^2 $ is the expected number of times the dark photon scatters as it crosses the halo and $V$ is the volume of the gas. The lengths
\begin{align}
    \ell_{\gamma_D}^C = \frac{ 3 m_X^2}{8\pi \alpha_D^2 n_X}; \quad \ell_{\gamma_D}^\mathrm{abs} = 3 \times 10^{-3} \frac{(m_X^9 T_X^5)^{1/2}}{\rho_X^2 \alpha_D^3},
\end{align}
are the dark photon Compton scattering length and absorption mean free path, respectively.

\begin{figure}[t]
\includegraphics[width=.49\textwidth]{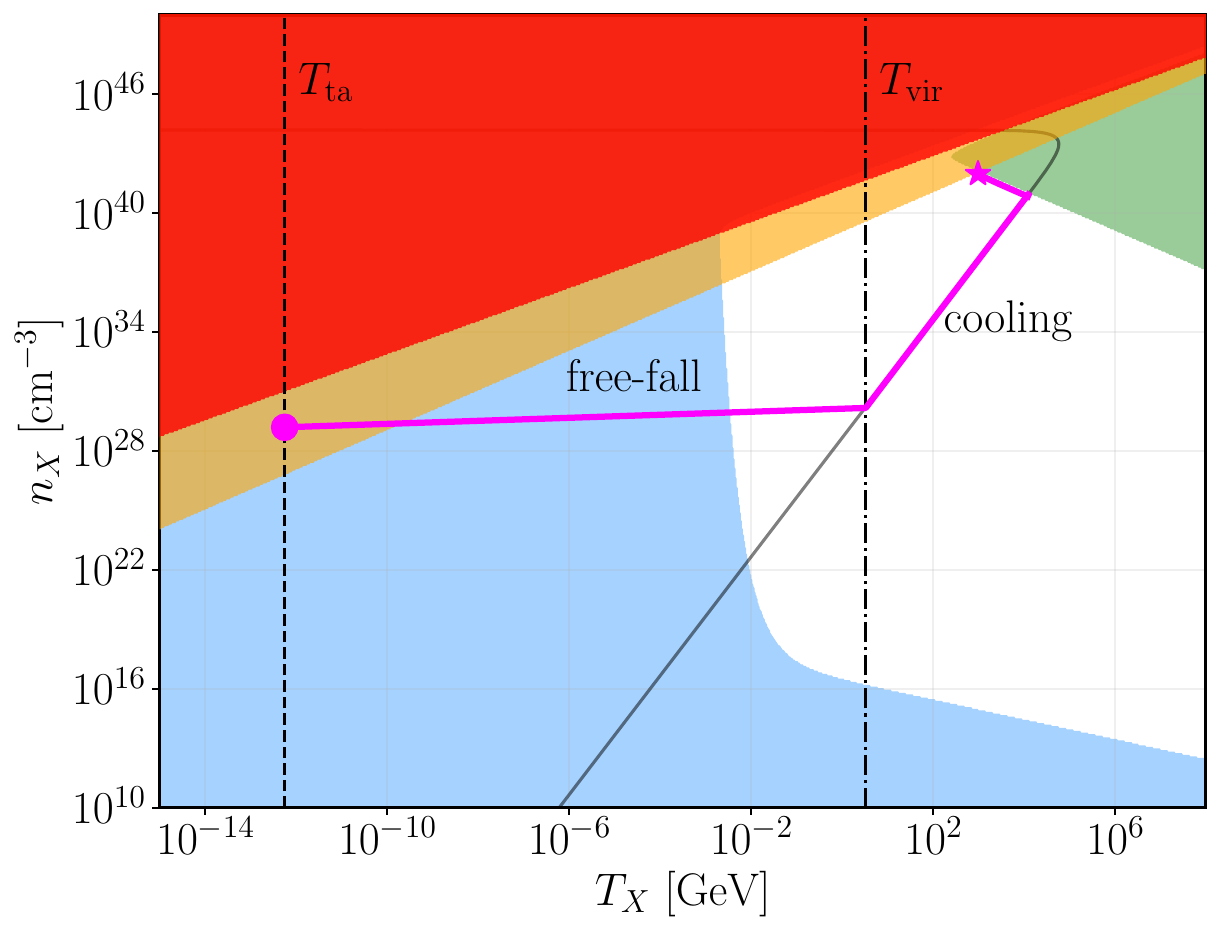}
\caption{\label{fig:traj} Collapse trajectory following turnaround for $m_X = 10^6$ GeV, $m_{\gamma_D} = 10^{-1}$ GeV, $\alpha_D = 0.1$, and $\delta_0 = 10^{-5}$. The blue shaded region on the left side of the plot is where $t_\mathrm{cool} > H_0^{-1}$ and so cooling is inefficient. The green shaded area in the top-right corner is where cooling is most efficient, $i.e.$ $t_\mathrm{cool} < t_\mathrm{ff}$. The red and orange regions in the upper-left portion indicate the stopping conditions of collapse, where the red region indicates when the gas is optically thick to bremsstrahlung and the orange area indicates when the gas becomes pressure-supported by the dark photon repulsive pressure. The solid gray curve indicates where nearly virialized contraction occurs, following a trajectory where $M_H=m_J$. The collapse trajectory is highlighted in pink. Nonlinear collapse proceeds starting with the point at turnaround ($\bullet$) with temperature $T_\mathrm{ta}$, indicated by the vertical dashed line. The gas of dark electrons collapses and heats up due to shock heating until it reaches half of its turnaround size, at which it virializes at a temperature of $T_\mathrm{vir}$, indicated by the dotted-dashed vertical line. It then proceeds in the nearly-virialized contraction and fragments once cooling becomes sufficiently efficient, stopping once it becomes pressure-supported ($\star$).}
\end{figure}

To model the nonlinear regime of structure formation, we follow the simplified, thermodynamical treatment carried out in Ref. \cite{Chang:2018bgx}. The dynamical evolution of the dark clump is given by

\begin{figure*}
\includegraphics[width=0.98\textwidth]{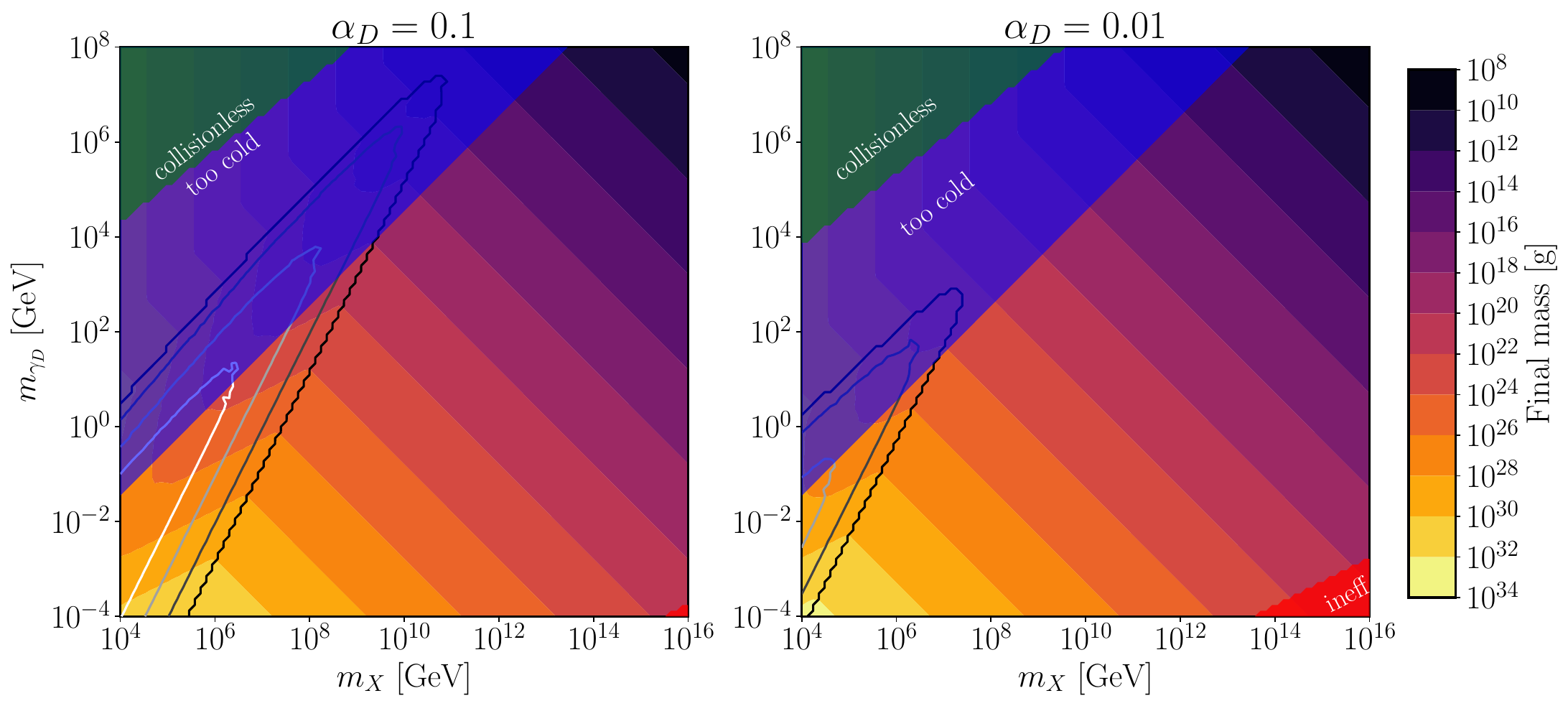}
\caption{\label{fig:dist} Distribution of final masses as a function of $m_X$ and $m_{\gamma_D}$ for $\alpha_D =0.1$ in the left panel and $\alpha_D = 0.01$ in the right panel. $\delta = 10^{-5}$ in both cases. The green region labelled ``collisionless'' is where the dark matter gas is not collisional at turnaround. The blue region labelled ``too cold'' is where the virial temperature is less than $m_{\gamma_D}$, and so bremsstrahlung cooling cannot occur. The red region labelled ``ineff'' indicates the region where the cooling time is longer than the age of the Universe. The solid white, light gray, dark gray, and black lines correspond to compactness values of $C = 10^{-3}, 10^{-2}, 10^{-1}$, and 0.5 respectively. All masses outside of the black contour lines do not undergo fragmentation, and are black holes. }
\end{figure*}

\begin{align}
    \frac{d \log T_{X} }{ d \log \rho_{X} } = \frac{2}{3} \frac{m_{X} P_{X} }{ \rho_{X} T_{X} }  - 2 \frac{t_\text{collapse} }{ t_\text{cool} },
\end{align}
where 
\begin{align}
    t_\text{collapse} = \left( \frac{d \log \rho_{X} }{dt} \right)^{-1}; \quad t_\text{cool} = \frac{3 T_{X}}{\Lambda },
\end{align}
and $t_\text{collapse}$ is set depending on the stage of collapse, which we discuss below. 

The collapse of primordial dark electron clouds can be separated into three stages. First, the $X$ overdensities will free-fall into virialized halos with radii about half of that of the overdensity at turnaround \cite{Bertschinger_1985}.  During free fall (as long as the gas is collisional) the gas will heat up rapidly due to shock heating.  Exchanging gravitational potential energy for thermal energy in the gas drives the temperature of the halo to about $T_\mathrm{vir} \sim m_X \delta_0/3$.  If $T_\mathrm{vir}>m_{\gamma_D}$, bremsstrahlung is efficient and the gas then undergoes a nearly-virialized contraction phase with $t_\mathrm{collapse} = t_\mathrm{cool}/6$. The gas collapses along a constant Jeans mass contour $m_J \sim M_H$ with $d \log T_{X}/ d \log n_{X} \simeq 1/3$. From here, the halo will continue contracting until either collapsing into a black hole or fragmenting into a collection of smaller pressure supported objects.  When $t_\mathrm{cool}>t_\mathrm{ff}$   for the entire collapse process, the halo collapses directly into a black hole.  Otherwise, if the temperatures and densities of the halos are large enough to make cooling efficient--as fast as collapse ($i.e.$ $t_\mathrm{collapse} = t_\mathrm{cool} \simeq t_\mathrm{ff}$), the gas fragments into smaller clumps. If cooling is very inefficient such that $t_\mathrm{cool}$ is larger than the age of the Universe, then the primordial halo today will be a pressure supported object with about the same temperature and density it had when it first virialized.

Once dark electron halos begin fragmenting, they will continue to do so until cooling becomes inefficient in one of two ways.  Either (i) the gas becoming optically thick when $\ell_{\gamma_D}^\mathrm{abs} = \sqrt{N_{\mathrm{sc}}} \lambda_J/2$, or (ii) the fragments become pressure-supported by the dark photon repulsive force. In the former case, once the gas becomes optically thick, we assume that the fragment will continue to cool due to surface cooling, and eventually become dark electron repulsion pressure-supported. The gas becomes repulsion pressure-supported when the second term in Eq. \eqref{eq:pressure} comes to dominate over the first. Once electron repulsion becomes the main source of pressure supporting the gas against collapse, cooling will not cause the ball to shrink any further as electron repulsion pressure is insensitive to the temperature of the gas. 

An example of a trajectory in which the gas fragments into DarkCOs is given in Fig. \ref{fig:traj}. The trajectory, indicated by the solid pink curve, starts from the point when the gas starts collapsing at turnaround and ends once fragmentation finishes. In this scenario, the stopping condition is due to the gas becoming pressure-supported by the second term in Eq. \eqref{eq:pressure}. 

Assuming that the compact objects evolve until they become pressure-supported, their compactness can be estimated as \cite{Chang:2018bgx}
\begin{align}
    C = \frac{GM}{r} \simeq \frac{m_X m_{\gamma_D} M}{\pi} \left( \frac{G^3}{\alpha_D} \right)^{1/2}. \label{eq:C_max}
\end{align}
A compactness of $C = 0.5$ corresponds to that of a black hole.

A possible landscape of resulting compact objects is given in Fig. \ref{fig:dist} for $\alpha_D = 0.1$ in the left panel and $\alpha_D = 0.01$ in the right panel. We find that the possible masses of final objects vary between $10^8$ g and $10^{34}$ g (between $10^{-25} M_\odot$ and $10^{1} M_\odot$) for dark electron masses between $10^4$ GeV and $10^{16}$ GeV and dark photon masses between $10^{-4}$ GeV and $10^8$ GeV. When the gas of $X$ particles cools efficiently enough to contract, but not efficiently enough to fragment, the halos collapse into black holes according to Eq. \eqref{eq:C_max}. For low values of $m_X$ and $m_{\gamma_D}$, cooling is more efficient, causing the halos to fragment instead of collapsing into a black hole.  This produces a range of DarkCOs.

One might worry that conservation of angular momentum either impedes collapse or causes the primordial halos to collapse into disks instead of spheres.  We estimate the typical angular momentum of these primordial halos at virialization by adding up the individual angular momenta of all of the dark electrons in a halo, assuming the magnitude of their velocity is set by $T_{\text{vir}}$, and the direction of their motion is random.  When we combine the individual angular momenta for the dark electrons as a random walk, we find that the typical angular momentum of these primordial halos at virialization always leads to centrifugal pressures in the halo that are small compared to the gas pressure during the entire collapse process. Thus, the angular momentum of the halos should not meaningfully alter their collapse.

\section{Pre-BBN phase transition}
\label{sec:inflation}

Since the Universe must be radiation-dominated at BBN for successful nucleosynthesis \cite{alphabetagamma}, the pre-BBN  EMDE that allowed dark electron structures to begin forming must come to an end $\textit{prior}$ to the start of BBN. Some additional dynamics are therefore required to revert from matter to radiation domination. Here we identify three possible ways to transition the Universe to SM radiation domination with $T_\mathrm{BBN} \approx {\rm MeV}$ at the onset of BBN:
\begin{enumerate}
    \item A thermal phase transition to an energy-dominated epoch dilutes the collapsed structures, restoring radiation domination in the Universe before BBN. This is motivated if the collapsed structures are associated with a heavy GUT sector \cite{Lyth:1995hj,Lyth:1995ka}.
    \item The collapsed structures have a substantial chemical potential or binding energy per $X$ particle. In this case, $X$ particles can be unstable to decay outside of dark structures, but stable against decay inside these structures. Provided that there is more energy density in $X$ particles outside of collapsed structures, this can lead to a radiation-dominated universe.
    \item The dark sector may consist of two particles, a dissipative component $X$ and a (perhaps heavier) particle $c$, which forms halos that $X$ collapses into. The heavy $c$ particle later decays to SM radiation. In this case, there needs to be a hierarchy between the energy density in $X$ and $c$, $\rho_{c} \gg \rho_{X}$.
\end{enumerate}
Other scenarios are certainly possible. For concreteness, we will focus on the first scenario of a thermal phase transition and leave the two other scenarios for future work. Suppose, as considered in Ref. \cite{Lyth:1995hj, Lyth:1995ka}, that a scalar field $\phi$ has a vacuum potential $V_0$ associated with a GUT scale symmetry breaking so that the potential of this scalar is given by
\begin{align}
    V(\phi) = V_0 - \frac{1}{2} m_\phi^2 |\phi|^2 + \lambda_s T^2 |\phi|^2 + \cdots, \label{eq:V_ns}
\end{align}
where $\lambda_s$ is the coupling of $\phi$ to some field in the thermal bath of particles. Note that higher order terms generated by coupling to heavy states ($e.g.$ $\frac{\phi^6}{\Lambda^2}$) stabilize $\phi$ in its potential in the limit $T\rightarrow 0 $ \cite{Lyth:1995hj}. For the potential above, there is a period of $V_0$-dominated thermal inflation, where $\phi$ is stalled by thermal bath friction from transitioning to the potential's minimum, while the final term of Eq.~\eqref{eq:V_ns} exceeds the second term. This period of thermal inflation begins when $V_0$ dominates over the dominant source of background energy, which in our scenario will be the dark matter density given by $\bar{\rho}_X = m_X \bar{n}_X$. Hence thermal inflation begins when $m_X \bar{n}_X \simeq V_0$, or in terms of the SM bath temperature,
\begin{align}
    T_i \sim \left( \frac{3 V_0}{g^*_\mathrm{MD} T_\mathrm{MD}} \right)^{1/3},
\end{align}
where $T_\mathrm{MD} \approx T_\mathrm{FO}/g_\mathrm{MD}^* \approx  m_X /(10 g_{\mathrm{MD}}^*)$ is the temperature where the Universe became matter-dominated, and so this determines the relative abundance of $X$ versus thermal bath particles. For brevity, we have denoted the number of effective relativistic degrees of freedom as $g^*_\mathrm{MD} = g_*(T_\mathrm{MD})$. 

Some comments are in order about the thermal bath fields to which $\phi$ is coupled. Generally, it is difficult to strongly couple $\phi$ to the same thermal bath particles that it would later decay into, since after thermal inflation, any fields it couples to strongly $\lambda_s = 1$ will obtain a large mass associated with the vev $\phi$ obtains in its zero temperature minimum (see the original model of \cite{Lyth:1995ka} and for an extended modern discussion see Section 5 of \cite{Dimopoulos:2019wew}). However, it is nevertheless possible to have the thermal stabilizing field be a field $\phi$ decays into, if $\lambda_S \ll 1$. For simplicity here we assume $\phi$ couples to a non-SM field with $\lambda_s = 1$, which becomes heavy at the end of thermal inflation, so that the inflaton then decays only to SM particles via a weaker coupling.

Thermal inflation will end when the temperature of thermal bath particles drops below the temperature at which the second-order correction terms in Eq. \eqref{eq:V_ns} are approximately equal, so that $T_\mathrm{PT} \approx m_\phi/\sqrt{2 \lambda_s}$. Once thermal inflation ends, we assume the decay of the scalar $\phi$ with potential energy $V_0$ into SM radiation happens quickly, and so the energy density of radiation after the decay is given by
\begin{align}
    \bar{\rho}_{r,\mathrm{RH}} = V_0 = \frac{\pi^2}{30} g^*_\mathrm{RH} T_\mathrm{RH}^4,
\end{align}
where $T_\mathrm{RH}$ is the reheating temperature after the decay of the scalar $\phi$. Since we have assumed the decay of $\phi$ happens quickly, the duration of the thermal inflationary period can be given in $e$-folds, by comparing the scale factors at the beginning of inflation and just at the end of inflation as $\mathcal{N} = \ln(a_\mathrm{PT}/a_i) $, so that
\begin{align}
    \mathcal{N} &= \ln \left( \frac{T_i}{T_\mathrm{PT}} \right) \approx \frac{1}{3} \ln \left( \frac{3V_0}{g^*_\mathrm{MD} T_\mathrm{MD} T_\mathrm{PT}^3 } \right) \nonumber \\
    & \simeq 10.3+\frac{1}{3}\ln \left( \frac{V_0}{10^{24}~{\rm GeV^4}} \cdot \frac{\rm GeV^3}{T_{\rm PT}^3} \cdot \frac{10^9~\rm GeV}{T_{\rm MD}} \cdot \frac{100}{g^*_{\rm MD}} \right)
    ,
\end{align}
where we have used the fact that $a \sim T^{-1}$. Small corrections to this estimate will arise depending on the coupling of $\phi$ to thermal bath fields \cite{Cho:2017zkj,Dimopoulos:2019wew}; in general we will utilize $\sim 10$ $e$-folds of thermal inflation expansion for our model regions of interest. Before returning to our dissipative dark sector scenario, we briefly note that it has recently been proposed that large fluctuations during thermal inflation might themselves result in PBHs \cite{Dimopoulos:2019wew,Bastero-Gil:2023sub}. However, we note that large fluctuations require $m_\phi /H \gtrsim 0.1$ during thermal inflation. In our case, $m_\phi \ll H$, and so we expect only a small change in the spectrum of sub-horizon perturbations, which will not be observable using standard cosmological methods \cite{Cho:2017zkj}.  

Working backwards from matter-radiation equality, we require that at reheating $\bar{\rho}_{X, \mathrm{RH}} \leq \bar{\rho}_{r, \mathrm{RH}} T_\mathrm{mr}/T_\mathrm{RH}$, where $T_\mathrm{mr}$ is the temperature at usual late-time matter-radiation equality. Hence we have that 
\begin{align}
    m_X &\lesssim 6\times 10^{15}~{\rm GeV}  \left(\frac{10}{g^*_\mathrm{PT}} \right) \left(\frac{g^*_\mathrm{RH}}{100} \right)^{1/4} \nonumber \\
    &~~\times \left(\frac{10^{-2}~{\rm GeV}}{T_\mathrm{PT}} \right)^{3} \left(\frac{V_0}{10^{24}~{\rm GeV^4}} \right)^{3/4}, \label{eq:m_X_upper}
\end{align}
where in the final inequality we have taken $T_\mathrm{mr} \simeq 0.8$ eV. We have also used the fact that the background density of $X$ particles following the phase transition is given by $\bar{\rho}_{X, \mathrm{RH}} = \pi^2 g^*_\mathrm{PT} T_\mathrm{PT}^3 T_\mathrm{FO}/30$, where again $T_\mathrm{FO} \approx m_X/10$ is the temperature at which the $X$ particles froze out and became non-relativistic. Note that the condition in Eq. \eqref{eq:m_X_upper} determines the relic abundance of the dark electrons. If we take a strict equality in Eq. \eqref{eq:m_X_upper}, then this would correspond with a scenario where the dark electrons constitute all of dark matter, as the contribution of the dark photons in the overall dark matter density are negligible.

Next we estimate a lower bound on the mass $m_X$, such that the density of $X$ particles in collapsed halos will exceed $V_0$ at the onset of thermal inflation. This will depend on the growth of the initial $X$ overdensity $\delta$, which will predominantly grow after the Universe become matter-dominated at $T_\mathrm{MD}$. At the time of $V_0$ domination, we require the overdensity grows to order one so that $\delta_{V_0} = \delta_0 a_{V_0}/a_\mathrm{MD} \gtrsim 1$, and so
\begin{align}
    m_X & \gtrsim 2 \times 10^{12}~{\rm GeV} \left( \frac{g^*_{\mathrm{MD}}}{100} \right)^{3/4} \nonumber \\
 &~~ \times \left( \frac{10^{-5}}{\delta_0} \right)^{3/4} \left( \frac{V_0}{10^{24}~{\rm GeV^4}} \right)^{1/4}.
\end{align}
Since we only strictly require that the period of thermal inflation occurs before BBN, so that $T_{PT} > 10~{\rm MeV}$, a wide range of thermal inflation energy densities is permitted, so that $V_0^{1/4} \sim 1-10^{10}~{\rm GeV}$.
Comparing the above equation to the lower bound presented prior, we find that if we fix the initial matter density contrast at $\delta_{0} = 10^{-5}$, a range of dark matter masses can be accommodated, $m_X \approx 10^{12} - 10^{18}$ GeV. However, within the parameters of the model presented here, collapsed objects formed from dark matter masses $m_X < 10^{12}$ GeV, might imply that $\delta_{0} > 10^{-5}$. Also, we note that our treatment has not been exhaustive. We have only required that $\delta \gtrsim 1$ at the time of thermal inflation as a reference point: a more detailed treatment of the growth of perturbations during a $V_0$ dominated phase would need to be undertaken to determine the exact condition for the overdensities to survive after thermal inflation. In addition, if the thermal inflationary period we have modeled with the simple potential given in Eq.~\eqref{eq:V_ns} lasted longer (in the case of a different potential with, $e.g.$ a first order phase transition), it should be possible to consider a cosmology with collapsed object formation and initial density perturbations that come to exceed unity well before the onset of a longer period of thermal inflation.

The fraction of dark electrons confined in dark structures depends on the primordial perturbation distribution, and on when the phase transition begins relative to when overdensities begin collapsing.  With the simplifying assumption that all dark electron density perturbations which have reached an overdensity of $\delta \gtrsim 1$, and thus decoupled from the background expansion of the Universe and become self gravitating, form dark electron halos that persist through the phase transition, and all overdensities that have not done so by the phase transition are stretched apart by thermal inflation, the fraction of dark electrons eventually ending up in compact structures is simply the fraction initially in perturbations that grow and begin collapsing prior to the onset of thermal inflation.   If, for simplicity, one assumes a Gaussian distribution of primordial density fluctuations with a standard deviation of $\delta_0\sim 10^{-5}$, and the phase transition occurs after density fluctuations of starting size $\sim 10^{-5}$ have had time to grow and begin collapsing, then one could estimate that all dark electrons in perturbations with starting densities at least one standard deviation above the average, corresponding to $\sim 17\%$ of dark electrons, would end up in collapsed structures. One can increase the fraction of dark electrons in compact structures by delaying the transition to radiation domination, allowing more halos time to form, or by assuming a larger initial density contrast, allowing halos to form more quickly.  Conversely, one can reduce the fraction of dark electrons in compact structures by allowing thermal inflation to start earlier, or by reducing the initial density contrast. Such adjustments do not alter the typical sizes of or evolution of the dark compact objects, just their abundance.

Note that this period of thermal inflation is advantageous because it allows for structure formation even after BBN. Given that the overdensities start to collapse in the EMDE, they have the lifetime of the Universe to continue their collapse since their densities will be always greater than the background density of the Universe (since $\rho_X > \bar{\rho}_{X,i} > V(\phi)$). Therefore the collapse will continue until they are pressure-supported, so long as the total collapse time is smaller than the age of the Universe. As alluded to in the previous section, this phase transition leaves the collapse of the black holes in Fig. \ref{fig:BHs} intact, leading to late-time formation of black holes.

\section{Phenomenological implications}
\label{sec:pheno}

This scenario has several phenomenological implications, stemming from both the exotic cosmology and the resulting landscape of collapsed objects. The EMDE and phase transition in the very early Universe can give rise to interesting phenomena such as the emission of GWs and modified structure formation \cite{Dalianis:2020gup, Ganjoo:2023fgg, Fernandez:2023ddy, Dalianis:2024kjr}.  Primordial halos can emit up to 0.3 $M_H$ in dark radiation as they collapse and cool.  This dark radiation can have observable effects by altering $N_\mathrm{eff}$ or delaying matter-radiation equality. Similar extended dark radiation emission scenarios have been proposed in the past as a way to solve the Hubble tension \cite{Pandey:2019plg}.  We save a more careful exploration of the ideas described above for future work and instead highlight the phenomena arising from the existence of dark compact objects below.

\subsection{Dark compact objects}

\begin{figure}
\includegraphics[width=.49\textwidth]{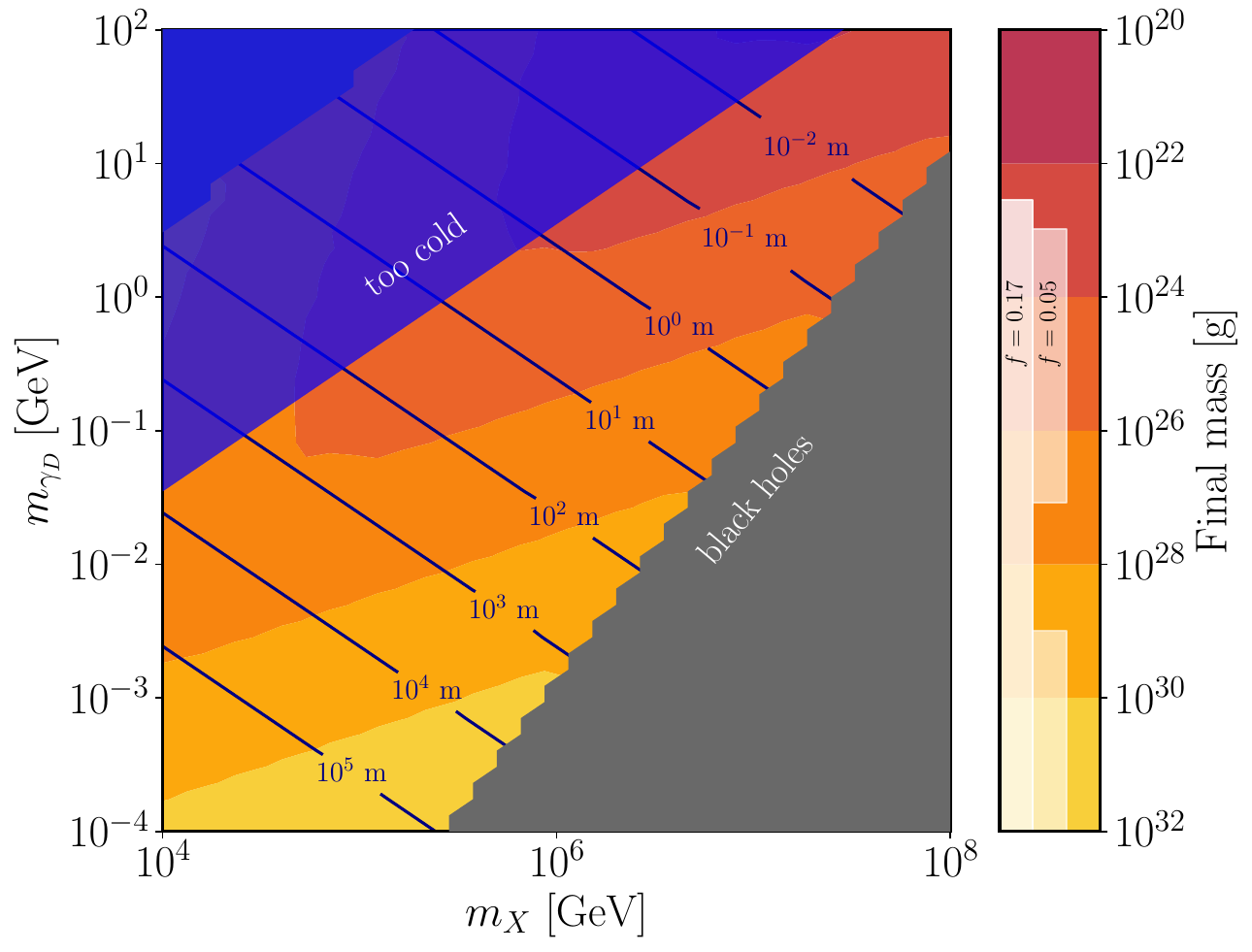}
\caption{\label{fig:fragments} A zoomed-in version of the left panel of Fig. \ref{fig:dist} for $m_X \leq 10^8$ GeV and $m_{\gamma_D} \leq 10^2$ GeV, focusing on the minimal fragments. The blue shaded region is where the virial temperature of the gas is too low for bremsstrahlung to occur. The gray region indicates where fragmentation does not occur, and final objects are black holes. Lines in navy are the characteristic radii of the fragments. The translucent white bars on the color bar indicate regions ruled out by previous microlensing searches \cite{bradley_j_kavanagh_2019_3538999} for a DarkCO fraction of $f=0.17$ on the left of the color bar and $f=0.05$ in the middle.}
\end{figure}

\begin{figure}
\includegraphics[width=.49\textwidth]{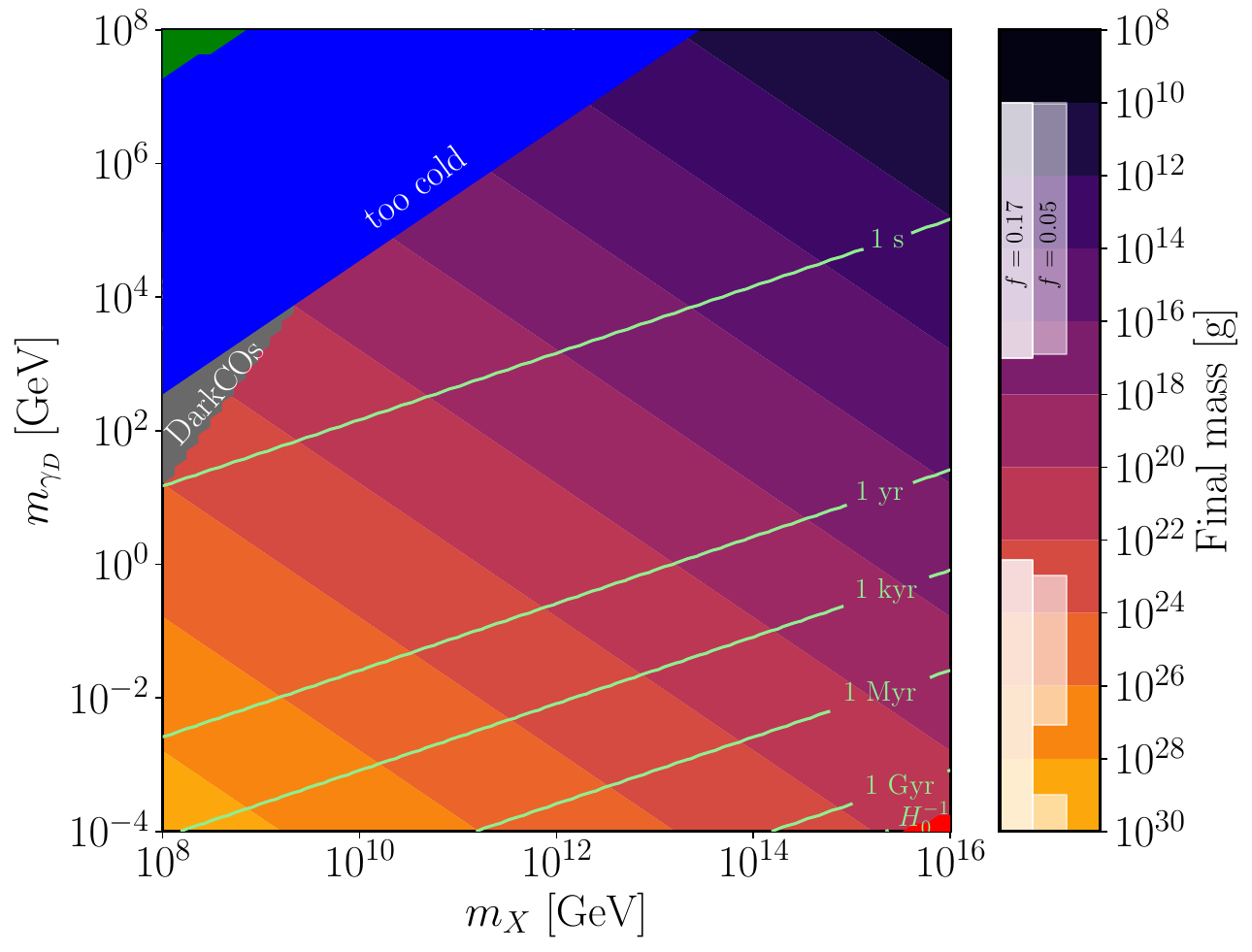}
\caption{\label{fig:BHs} A zoomed-in version of the left panel of Fig. \ref{fig:dist} for $m_X \geq 10^8$ GeV, focusing on the black holes. The green, blue, and red shaded regions are the same as Fig. \ref{fig:dist}, $i.e.$ the collisionless gas regime, the too cold regime, and the ineffective cooling regime, respectively. The gray shaded area is the region where fragmentation occurs, and the final objects do not form black holes. Curves in light green are the collapse timescales at virialization. The translucent white bars on the color bar indicate regions ruled out by previous MACHO searches for a PBH fraction of $f=0.17$ on the left of the color bar and $f=0.05$ in the middle. Digitized PBH constraints obtained using  \cite{bradley_j_kavanagh_2019_3538999} and \cite{Carr:2020gox}. }
\end{figure}

As discussed in Sec. \ref{sec:final_masses}, our mechanism produces dark compact objects of varying sizes and masses depending on the model parameters. We classify these as either DarkCOs or black holes. Zooming in on the region of parameter space in Fig. \ref{fig:dist}, Fig. \ref{fig:fragments} shows possible masses for the fragmented DarkCOs along with their characteristic radii. We find that these final fragments get larger in size as one goes to lower values of $m_X$ and $m_{\gamma_D}$, since the dark photon repulsive force in Eq. \eqref{eq:cs_sq} comes to dominate for lower densities, leading to an increased Jeans length. 

These DarkCOs are very compact objects and can thus be treated as point sources -- making them susceptible to microlensing constraints. Indeed, we can apply existing constraints on MACHOs~\cite{bradley_j_kavanagh_2019_3538999} based on data from microlensing surveys such as Subaru/HSC \cite{Niikura:2017zjd}, MACHO \cite{MACHO:2000qbb}, EROS-2 \cite{EROS-2:2006ryy}, OGLE-III+IV \cite{Niikura:2019kqi, Mroz:2024mse, Mroz:2024wag}, and Kepler \cite{Griest:2013aaa}, to the DarkCOs predicted in this work. The overlaid white bars in the color bar of Fig. \ref{fig:fragments} indicate ruled-out DarkCO mass ranges assuming a fraction $f$ of dark matter is in these compact objects. We have considered $f=0.17$ and $f=0.05$ corresponding to the assumption that dark electrons comprise all of the dark matter and that $17\%$ or $5\%$ of the dark electrons have formed compact structures. These fractions correspond to scenarios in which thermal inflation begins after all density perturbations with an initial overdensity at least $1\sigma$ and $2\sigma$ above the background density have had time to collapse. For a fraction of $f=0.17$, only lower mass DarkCOs (very top right of Fig. \ref{fig:fragments}) are permitted due to the strong constraint coming from Subaru/HSC, OGLE-III+IV, and EROS-2. However, for a lower value of $f=0.05$, the constraint from EROS-2 is weaker, and hence an additional window opens up between $10^{27}$ g (coming from Subaru/HSC \cite{Niikura:2017zjd}) and $10^{29} $ g (from the 20 year dataset of OGLE-III+IV \cite{Mroz:2024mse, Mroz:2024wag}). 

We have treated the DarkCOs as point sources due to their high density and compactness, and checked that their characteristic radii are smaller than their Einstein radii. However, for larger extended dark objects, their non-trivial structure can result in weakened microlensing constraints \cite{Croon:2020wpr, Croon:2020ouk}.  Indeed, Ref.~\cite{Bai:2020jfm} showed that extended compact objects in our DarkCO mass range can make up all of the dark matter, if they have large enough radii.  While the DarkCOs presented in the parameter space we consider are too small in mass to be found in current microlensing surveys, it is possible that mergers of the DarkCOs or accretion may further increase their sizes. We leave a more careful treatment of the mass and radius distributions of merged DarkCOs and their resulting constraints to future work. CMB measurements can also be used to constrain the abundance of DarkCOs \cite{Croon:2024rmw}, but only at masses somewhat above the mass range of the model parameters we have considered.

\subsection{Black Holes}
Meanwhile, Fig.~\ref{fig:BHs} shows a zoomed-in version of the predicted black hole parameter space in Fig.~\ref{fig:dist}, with collapse timescales at virialization overlaid on top. Similar to Fig.~\ref{fig:fragments}, we have indicated allowed PBH masses assuming a PBH fraction of $f=0.17$ and $f=0.05$. The lower mass exclusion band arises from evaporation constraints (see \cite{Carr:2020gox} for a thorough review), while the same microlensing constraints discussed for Fig.~\ref{fig:fragments} apply for higher masses. PBHs with masses $M \lesssim 10^{10}$ g are unconstrained because they decay too early to impact BBN. The allowed asteroid-mass window between the two constrained regions for both values of $f$ arise from the gap between evaporation constraints and microlensing constraints. Recently, Ref.~\cite{Agius:2024ecw} reexamined CMB constraints on primordial black holes to set strong limits on the PBH abundance, but only for masses somewhat above the largest masses we consider.

The collapse timescales increase as the dark photon masses decrease, primarily since the initial density of the clump decreases as a function of the dark photon mass. Interestingly, objects with cooling times comparable to the age of the Universe at BBN imply the late-time formation of PBHs, with some parameter space having black holes forming today or in the future. Late-forming PBHs have been studied in previous works, though these relied on a different PBH production mechanism \cite{Chakraborty:2022mwu, Lu:2022jnp, Picker:2023ybp}.

\begin{figure*}
\includegraphics[width=.98\textwidth]{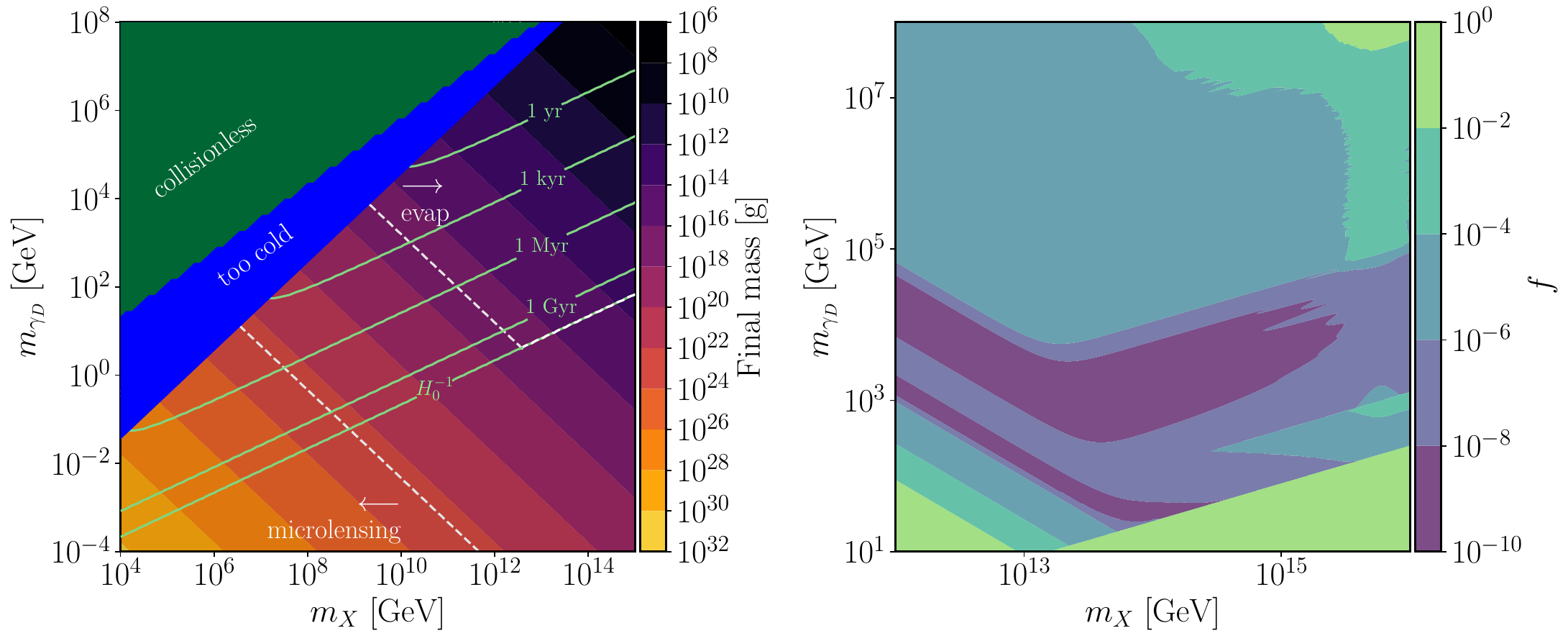}
\caption{\label{fig:M_and_f} \textbf{Left:} Distribution of final masses with $\alpha_D = 10^{-7}$. The green and blue regions are as described in Fig. \ref{fig:dist}. Curves in light green are the collapse timescales at virialization. The regions indicated by white dashed lines show where PBH constraints are relevant for a PBH fraction of $f = 0.17$. \textbf{Right:} Evaporation constraints for the fraction $f$ of dark matter in late-forming PBHs assuming $\alpha_D = 10^{-7}$. }
\end{figure*}

However, decreasing $\alpha_D$ modifies the existing PBH bounds in multiple ways. First, it makes cooling less efficient, causing the halos to collapse into PBHs at later times. As shown in the left panel of Fig.~\ref{fig:M_and_f}, for some regions of the dark electron parameter space, reducing $\alpha_D $ to $ 10^{-7}$ leads to the formation of low mass black holes which take longer to collapse than to decay. As the evaporation of these late forming PBHs would happen at different times than usually expected for PBHs, evaporation constraints are altered. The majority of the parameter space in the lower right corner of the left panel of Fig.~\ref{fig:M_and_f} is unconstrained since these dark halos would not have collapsed into black holes yet.  

Second, when $\alpha_D$ is reduced to $10^{-7}$, the PBHs generated from the heaviest dark electrons will be near-extremal (as defined below) when they form. This causes them to have a cooler horizon temperature, radiate a smaller fraction of their mass, and emit radiation more slowly than uncharged black holes of the same mass. The modified constraints on late forming and near-extremal decaying black holes are shown in the right panel of Fig.~\ref{fig:M_and_f}.  The constraints presented here come from modifying existing evaporating black holes limits based on observations of primordial elemental abundances from BBN \cite{Keith_2020}, of CMB anisotropies \cite{Poulin_2017}, and from gamma-ray emission at late times \cite{Carr_2010}.

When describing the black holes as near-extremal, we are referencing the fact that black holes formed when a dark electron cloud collapses will retain the charge of the dark electrons that formed it.  Such charged black holes cannot evaporate away to arbitrarily small masses as is often predicted for Schwarzschild black holes, but will instead evaporate down to an extremal mass, dictated by the the black hole's starting mass, $M_\mathrm{BH}$, as well as $m_X$, $m_{\gamma_D}$ and $\alpha_D$.  If the mass of the black hole is far from its extremal limit, the event horizon temperature and radius will resemble those of a Schwarzschild black hole.  When the mass comes within an order of magnitude of the extremal limit, becoming ``near-extremal'', its behavior will deviate significantly from that of a Schwarzschild black hole.  As it approaches its extremal mass, the event horizon radius will stop shrinking and the horizon temperature will approach zero. See Appendix~\ref{appendix:charged_BHs} for details. Note that when $\alpha_D=0.1$, none of the black holes predicted in this work will be near-extremal, leaving existing black hole decay constraints unchanged.

The constraints on late forming and near-extremal black holes when $\alpha_D=10^{-7}$ from BBN \cite{Keith_2020}, of CMB anisotropies \cite{Poulin_2017}, and from gamma-ray emission at late times \cite{Carr_2010} are presented in the right panel of Fig. \ref{fig:M_and_f}.  We make the simplifying assumption that the black hole energy is deposited mostly at a time $t\propto t_{\text{cool}}+t_{\text{decay}}$, where $t_{\text{decay}}$ is the time it takes the black hole to decay. For charged black holes $t_{\text{decay}}$ must be calculated numerically.  In Figure \ref{fig:M_and_f}, these constraints largely follow curves of constant $t_{\text{cool}}+t_{\text{decay}}$,  as the decay bounds usually depend on the timing of energy injections. The constraints disappear when $t_{\text{cool}}$ is larger than the age of the Universe, as these black holes have yet to form.  The constraints consistently weaken for the highest values of $m_X$.  This is because the black holes formed from heavier dark electrons tend to be closer to their extremal mass, and will emit a smaller fraction of their total mass ($\sim 10\%$) before becoming extremal.  We give more detailed descriptions of how we calculated these constraints in Appendix \ref{appendix:B}.

Black holes that collapse quickly and take longer than the lifetime of the universe to decay happen to be in a mass range where the charge of the black hole does not alter the observed radiation today.  These are subject to the usual radiation constraints for black holes with lifetimes older than the Universe \cite{Carr_2010}.

While we have estimated the constraints on charged late forming black holes for reference, these constraints generally rely on the entire emission history of the black holes.  One would have to consider these black holes' unique evolution history (rather than taking averages), to get precise constraints. Recently, a numerical analysis of constraints on the PBH abundance from BBN was studied in Ref. \cite{Boccia:2024nly}, which placed the strongest bounds on the PBH mass between $10^8$ g and $10^9$ g. An interesting extension would be to study how these bounds differ in the case of the delayed black holes presented in this work.   We leave these more careful treatments to future work.

Furthermore, a particularly interesting and novel observable feature of this scenario is the late-time flashes from rapid evaporation of light PBHs \cite{Hawking:1974rv, Boluna:2023jlo} which can be searched for in gamma ray detectors. While the charges on these black holes will prevent them from decaying to arbitrarily low masses (high temperatures), they can still get hot enough to probe energies inaccessible to colliders on Earth. Late-time forming black holes with light masses have been suggested as a possible solution to antiproton, gamma-ray, and Helium excesses from the galactic center \cite{Picker:2023lup, Korwar:2024ofe}.    

Finally, in addition to the GW emission from the exotic cosmology, binary mergers of both the compact objects and black holes produced in this scenario would give rise to GW emissions \cite{Nakamura:1997sm, Raidal:2017mfl, Shandera:2018xkn, Singh:2020wiq, Diamond:2021dth}. Furthermore, the evaporation of PBHs can also affect the background of stochastic GWs \cite{Anantua:2008am, Ireland:2023avg}. We leave these considerations for future studies. We comment that there are also dynamical constraints on compact objects, such as the disruption of dwarf galaxies via dynamical heating \cite{Brandt:2016aco, Graham:2023unf} and wide binaries \cite{Yoo:2003fr, Monroy_Rodr_guez_2014}. However, these only become relevant at compact object masses larger than $\sim 10^{34}$ g, and are weaker than microlensing constraints until $\sim 10^{36}$ g, which is larger than any of the final object masses we have considered in this work. Other dynamical effects, such as the accumulation of compact objects in the galactic nucleus via dynamical friction \cite{Carr_1999} can also place constraints. Once again however, these become relevant at compact object masses higher than the ones in this work.

\subsection{SIDM restrictions}

In the scenario we consider, not all of the dark matter ends up in gravitationally collapsed structures by the present day. Assuming that an $\mathcal{O}$(1) fraction of dark matter is not bound in collapsed structures, we can ask whether the self interaction generated by the dark photon would have observational consequences, as limits have been set on dark matter self interactions from observations of galaxies and galaxy clusters. The most famous of these bounds comes from observations of the Bullet Cluster, first computed in Ref.~\cite{Markevitch:2003at}. The Bullet Cluster is actually a system of two merging clusters, and constraints on dark matter self interactions can be set based on the offset between the gas and dark matter distributions, the velocity of the smaller subcluster, and the very survival of that subcluster. Assuming that dark matter scatters isotropically, which is true for a contact interaction---i.e. when the mediator mass is large compared to the momentum transfer---this last consideration results in a constraint on the dark matter self interaction cross section of approximately $\sigma < 1$ cm$^2$/g.

At the velocities typical of the Bullet Cluster, the condition of a contact interaction is satisfied in the upper left region of Fig.~\ref{fig:dist}, specifically the green and part of the blue shaded region. We compute the dark matter self interaction cross section per unit mass throughout this region, and find it to always be much less than $10^{-10}$ cm$^2$/g, easily evading the Bullet Cluster constraint.

In most of the parameter space of Fig.~\ref{fig:dist}, however, the mediator mass is light compared to the dark matter momentum, and the above limit cannot be naively applied. However, the limit can be easily recomputed for an arbitrary scattering angle distribution. The limit from the survival of the subcluster is based on the requirement that less than 30\% of the dark matter in the subcluster be upscattered beyond its escape velocity. Scanning over the entire parameter space of Fig.~\ref{fig:dist}, we find that the fraction of dark matter upscattered past escape velocity is at most $10^{-10}$, once again many orders of magnitude smaller than the observed limit.

Finally, we consider the limit on self interactions set by Ref.~\cite{Pardo:2019wie}, based on the warping of galactic disks. They define a reference cross section $\tilde{\sigma} = 16\pi \alpha_D^2/m_X^2$, which they constrain to be $\tilde{\sigma}/m_X < 3 \times 10^{-13}$ cm$^2$/g. Scanning over all of our parameter space that does \emph{not} lie in the contact regime, we find a maximum value of $\tilde{\sigma}/m_X \simeq 10^{-16}$ cm$^2$/g. This is not a difference of 10 orders of magnitude, as it was with the Bullet Cluster constraints, but our model is still easily allowed in all the parameter space we consider.

\section{Conclusions}
\label{sec:conclusion}

In this work we have demonstrated how a simple, dissipative dark sector could fall out of equilibrium with the visible sector in the very early Universe and be responsible for an early period of matter domination before BBN -- leading to the formation and collapse of DarkCOs and primordial black holes.  This offers a new production mechanism for compact objects in the dark sector, and provides a concrete scenario where dissipative compact object formation can occur in the dark sector, without restricting this dissipative dynamics to occur for only a subcomponent of dark matter. In the scenario we have outlined, the usual bounds on dark matter self-interactions do not apply, since dissipation and strong self-interactions are only at play in the very early Universe.

Interestingly, for some parameter space this production mechanism  predicts the delayed formation of PBHs. These delayed PBHs form at late times, modifying existing constraints on PBHs while providing additional novel signatures that can be searched for via GWs and late-time radiation emission. 

There are additional avenues open to future investigation, especially with regards to the late time behavior of compact objects and black holes. In particular, as shown in Figs.  \ref{fig:dist} and \ref{fig:fragments}, there is a substantial population of DarkCOs predicted for dark matter masses $m_X \approx 10^4-10^8$ GeV, which will cool and emit $\gamma_D$ particles. While in this work we have assumed the dark photon $\gamma_D$ has no coupling to SM particles, it will be interesting to consider whether there is detectable, diffuse radiation from these objects if $\gamma_D$ couples to the SM through, $e.g.$ a vector portal operator. We leave further investigation of these and other rich phenomena associated with DarkCOs and late-forming PBHs to future work.

\begin{acknowledgments}
We thank Daniel Ega\~na-Ugrinovic, Himanish Ganjoo, James Gurian, Felix Kahlhoefer, Bart Ripperda, and Sarah Sadavoy for useful discussions. The authors and this work were supported by the Arthur B. McDonald Canadian Astroparticle Physics Research Institute, the Natural Sciences and Engineering Research Council of Canada (NSERC), and the Canada Foundation for Innovation. Research at Perimeter Institute is supported by the Government of Canada through the Department of Innovation, Science, and Economic Development, and by the Province of Ontario. JLK is supported by an NSERC CGS-D. CVC was also generously supported by Washington University in St. Louis through the Edwin Thompson Jaynes Postdoctoral Fellowship.
\end{acknowledgments}

\onecolumngrid
\appendix
\section{Black holes with a broken U(1) charge} 
\label{appendix:charged_BHs}
A black hole made up of a collection of dark electrons may carry the same dark charge as the sum of the electrons that comprise it.  Because the dark photon is heavy, the dark electrons, and by extension the black hole they form, will have a Yukawa-like dark electric potential $ V \propto \alpha_D n_{\chi} e^{m_{\gamma_D}r}/r$, where $r$ is the distance from the black hole.  The properties of such black holes are worked out in Ref.~\cite{2019IJMPD..2850120M} and are presented here for convenience. 

Such black holes are expected to be spherically symmetric and have a metric of the form
\begin{equation}
    ds^2 = -f(r) dt^2 + \frac{dr^2}{f(r)}+r^2(d\theta^2+\sin^2\theta d\phi^2),
\end{equation}
where 
\begin{align}
     f(r) &= 1-\frac{G M}{r} +G \frac{m_{\gamma_D}^4 r^2 Q^2}{6}\mathcal{E}_1(m_{\gamma_D}r)  -\frac{G Q^2(m_{\gamma_D}^3 r^3-m_{\gamma_D}^2r^2+2 m_{\gamma_D}r-6)e^{-m_{\gamma_D}r}}{6r^2} \label{eq:f_r} .
\end{align}  
Here, $M$ is the mass of the black hole, and $Q$ is its charge. For black holes composed of dark electrons, $Q = \sqrt{4\pi\alpha_D}M/m_X$. The quantity $\mathcal{E}_1(x)$ is given by
\begin{align}
    \mathcal{E}_1(x) = \int_0^{\infty}\frac{e^{-xt}}{t}dt.
\end{align}

The horizons of the black hole are located at the values of $r$ that satisfy $f(r)=0$.  There may be two horizons, similar to a standard charged black hole, one horizon, meaning the black hole is extremal and has zero surface temperature, or no horizons, meaning the black hole is super-extremal and would not have been able to form in the first place.  There is no simple analytical solution for the horizon locations, and these must be calculated numerically.

A black hole will Hawking radiate from the outer horizon, located at $r_h$, with a temperature set by
\begin{align}
    T=\frac{f'(r)}{4\pi}\Big\rvert_{r_h}.
\end{align}
When $r_h\gg m_{\gamma_D}^{-1}$, the contribution from the charge terms in Eq. \eqref{eq:f_r} is exponentially suppressed, and the event horizon sits very near $r_h \sim 2GM$ as expected for an uncharged black hole.  The temperature will also be very close to $T \sim (8\pi GM)^{-1}$, again resembling an uncharged black hole. As the black hole loses mass through Hawking radiation, $r_h$ will shrink, until $r_h\gtrsim m_{\gamma}^{-1}$, and the charged terms are no longer exponentially supressed.  The surface temperature of the black hole then drops, causing the evaporation rate to slow down as the black hole approaches its final extremal radius and mass. Note that it will technically take an infinite amount of time for the black hole to reach its extremal state as the surface temperature tends toward zero when the black hole approaches extremality.  

\section{Modification of evaporation constraints}
\label{appendix:B}
\subsection{BBN constraints}
The relevant BBN constraints on evaporating black holes come from observations of the He$_4$/He$_3$ and He$_4$/D ratios.  Early injections of energetic particles can dissociate Helium nuclei and alter these ratios.   When the temperature of the standard model bath is $T\gtrsim 0.4$ keV, injected hadrons and mesons are most efficient at dissociating nuclei (hadrodissociation), while injected electromagnetic particles are most efficient at lower temperatures (phtotodissociation).  We determine these constraints following the approach of Ref.~\cite{Keith_2020}, by modifying limits on dark matter that decays into pairs of quarks presented in Ref.~\cite{Kawasaki_2018}.  
    
We assume that the black holes will yield approximately the same constraints as an equivalent distribution of decaying particles, if both objects decay to similar distributions of particles in the early universe at about the same time.  
For black holes decaying during the photodissociation period, we calculated the average energy of quarks and electromagnetic particles emitted over its lifetime $\langle E \rangle$ and matched the black hole with a decaying particle with $m=2\langle E \rangle$.  During the hadrodissociation era, the constraints are sensitive to the number of energetic hadrons emitted from the black hole rather than their energy \cite{Keith_2020}.  For these black holes, we calculated the average energy for quarks to produce a hadron $\langle E_q \rangle$, along with the number of hadrons produced by the black hole, which scales as $T_\mathrm{BH}^{3.3}$ \cite{Keith_2020}, and matched them to decaying particles with $m=2\langle E_q \rangle$. Next, we then matched the black hole lifetimes to decaying particle lifetimes.  

During the photodissociation era, we take the average particle injection time to be $t_\mathrm{collapse}$ plus the time needed for the black hole to radiate half of the mass it will lose before becoming extremal, $t_\mathrm{half}$, which must be calculated numerically. During the hadrodissociation era, we match the median hadron injection time, $t_\mathrm{collapse}$ plus the time at which the black hole will have injected half of the hadrons it is going to emit, to a decaying particle with the same median injection time.  For a particle with a lifetime $\tau$, the median injection time is $\tau \ln(2)$.  As the black holes will become extremal before radiating away all of their mass, we divide the matched constraints on the decaying particles by the fraction of black hole mass that can be radiated away before the black hole becomes extremal, $f_\mathrm{loss}$, to get the final black hole constraints.

\subsection{CMB constraints}
We used existing CMB constraints on heavy particles decaying to electromagnetic radiation \cite{Poulin_2017} to estimate the CMB limits on radiation from late forming black holes.  Here we matched the black holes to decaying particles with similar particle injection times. We take the limits on the amount of radiation produced by a decaying particle with lifetime $\tau$ and apply it to the black holes with $t_\mathrm{collapse}+t_\mathrm{half}=\tau$.  These constraints are also divided by $f_\mathrm{loss}$ as the black holes will not fully evaporate away.  These constraints break down once the black hole formation plus radiation time becomes long enough, $t_\mathrm{collapse}+t_\mathrm{half}\gtrsim 1.5\times 10^{16}$ s, which is around the time of reionization.  Black holes that inject energy after reionization will do little to alter the ionization history or CMB observables.

\subsection{Gamma-ray constraints}
Black holes that collapse and decay before the present are subject to constraints from gamma-ray emission, as explored in Ref.~\cite{Carr_2010}.  The constraints come from black holes emitting quarks and gluons which confine into mesons, which themselves further decay into gamma-rays that would be observable today.  The number of photons produced at the spectral peak of $m_{\pi}/2$ scales linearly with the rate of primary particle emission per unit energy, $\tfrac{d^2N}{dEdt}\Big|_{E\sim T}$, evaluated at $E\sim T$, and the amount of time the black hole can radiate for. The peak energy of the observable signal today will be redshifted by $(1+z_\mathrm{inj})^{-1}$, where $z_\mathrm{inj}$ is the redshift at which the energy was injected.  For our delayed black holes, we take this to be $z_\mathrm{inj} = z(t_\mathrm{collapse}+t_\mathrm{half})$. The resulting limit that emerges from rescaling the limits presented in Ref.~\cite{Carr_2010} is
\begin{align}
    f&<2\times 10^{-9}\left(\frac{M_\mathrm{BH}}{5\times 10^{14}\text{g}}\right)^{-1}(1+z_\mathrm{inj})^{1.4}  \times \frac{\int_0^{t^*}\left(\frac{d^2N}{dEdt}\right)^s\Big|_{E\sim T} dt} {\int_0^{\infty}\left(\frac{d^2N}{dEdt}\right)^c\Big|_{E\sim T} dt},
\end{align}
where $\left(\frac{d^2N}{dEdt}\right)^s$ denotes the particle flux from a Schwarzschild black hole of starting mass $M_\mathrm{BH}$ and corresponding lifetime $t^*$, and $\left(\frac{d^2N}{dEdt}\right)^c$ is the flux from a charged black hole made of dark electrons with staring mass $M_\mathrm{BH}$.  The upper time limit on the integral is infinite as these black holes never quite reach their extremal mass.

\twocolumngrid
\bibliography{refs}

\end{document}